# Micro-engineered $CH_3NH_3PbI_3$ nanowire/graphene phototransistor for low intensity light detection at room temperature

*M. Spina, M. Lehmann, B. Náfrádi , L. Bernard, E. Bonvin, R. Gaál, A. Magrez [a], L. Forró, E. Horváth**

Laboratory of Physics of Complex Matter (LPMC), Ecole Polytechnique Fédérale de Lausanne, 1015 Lausanne, Switzerland

[a] Crystal Growth Facility, Ecole Polytechnique Fédérale de Lausanne, 1015 Lausanne, Switzerland

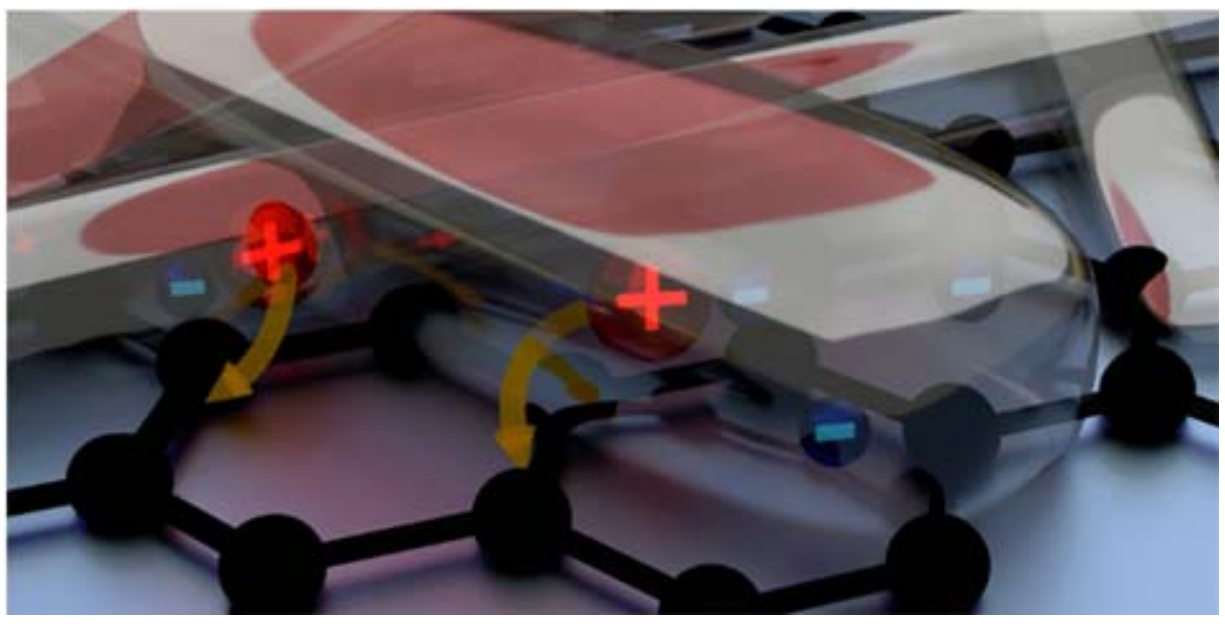

**Methylammonium lead iodide perovskite has revolutionized the field of third generation solid-state solar cells leading to simple solar cell structures[1] and certified efficiencies up to 20.1%[2, 3]. Recently the peculiar light harvesting properties of organometal halide perovskites have been exploited in photodetectors where responsivities of ~3.5 A/W and 180 A/W have been respectively achieved for pure perovskite-based devices[4-6] and hybrid nanostructures[7]. Here, we report on the first hybrid phototransistors where the performance of a network of photoactive Methylammonium Lead Iodide nanowires (hereafter $MAPbI_3NW$) are enhanced by CVD-grown monolayer graphene. These**

**devices show responsivities as high as ~2.6x10$^6$ A/W in the visible range showing potential as room-temperature single-electron detector.**

Certain characteristics (*i.e.* direct bandgap[8], large absorption coefficient[9], long charge diffusion lengths[10, 11]) make $MAPbI_3$ appealing for several optoelectronic applications. In optoelectronic devices, it was first introduced as sensitizer in dye-sensitized solar cells (DSSCs) by Miyasaka and coworkers[12]. In the following years, solid-state perovskite-based solar cells have risen to the forefront of thin film photovoltaics research, recently reaching certified efficiencies of 20.1%.[2, 3] Despite the intense research in the field of photovoltaics the potential use of $MAPbI_3$ in photodetectors is still largely unexplored. [4,7] This is partly due to the relatively high intrinsic resistivity and thus low photoelectron collection efficiency of $MAPbI_3$. This shortcoming of $MAPbI_3$ was overcome by X. Hu *et al.*[4] by depositing a pure film of $MAPbI_3$ nanoparticles on a flexible ITO/PET substrate. Accordingly responsivities as high as ~3.5 A/W have been attained. An additional 50-fold improvement has been shown by replacing ITO/PET substrate by graphene in hybrid $MAPbI_3$/graphene photodetectors by Y. Lee and coworkers[7]. This increase is partially attributed to the superior electronic characteristics of graphene over ITO/PET, but more importantly to a photo-doping/gating effect[7, 13].

The electronic performance of graphene, however, is known for its sensitivity to chemical alterations. In particular the carrier mobility in graphene changes by orders of magnitude by the random surface potential enforced by the substrate and by the coating layer. In order to unravel the true photodetection potential of $MAPbI_3$/graphene heterostructures a precise control of the $MAPbI_3$-graphene contacts are required. To this end, we have micro-engineered devices where a network of $MAPbI_3$ *nanowires* ($MAPbI_3NW$) has been combined with CVD-grown monolayer graphene. The resulted phototransistors showed responsivities as high as $2.6x10^6$ A/W *i.e.* a value 4 orders of magnitude higher than earlier literature reports.

A $MAPbI_3$ nanowire/graphene phototransistor was constructed from a network of perovskite nanowires deposited by the slip-coating method reported by Horváth *et al.*[14] onto single layer graphene field-effect transistor (FET). (Detailed description of the graphene-FET fabrication is given in the supporting information.) A combined optical and scanning electron microscope (SEM) micrograph of a representative device is shown in Figure 1a. Raman analysis of the graphene layer shows a threefold increase of the D-peak ($I_D$) over G-peak ($I_G$) ratio before and after covering the graphene with a network of perovskite nanowires (Figure 1b). This may imply the formation of C-I bonds on the as-grown graphene[15]. These additional scattering centers significantly modify the transfer characteristics of the devices (Figure 1c). The pristine graphene FET showed p-doped behavior with a charge neutrality point (Dirac point) above 50 V. This is attributed to the reactions of defects and PMMA residues with p-doping chemical vapor adsorbents such as $H_2O$ and CO[16, 17]. As depicted in Figure 1d the Dirac point of the single layer graphene transistor shifted heavily towards more positive voltages after the slip-coating of $MAPbI_3$ nanowires from DMF solution indicating a significant chemical doping of graphene by $MAPbI_3$. Additionally, the charge carrier mobility of graphene was lowered by a factor of 10 as a consequence of $MAPbI_3NW$ deposition indicating the appearance of significant random potential at atomic length-scales on the $MAPbI_3NW$/graphene interface. As we will show both of these two consequences of the $MAPbI_3NW$/graphene hybridization decrease the responsivity.

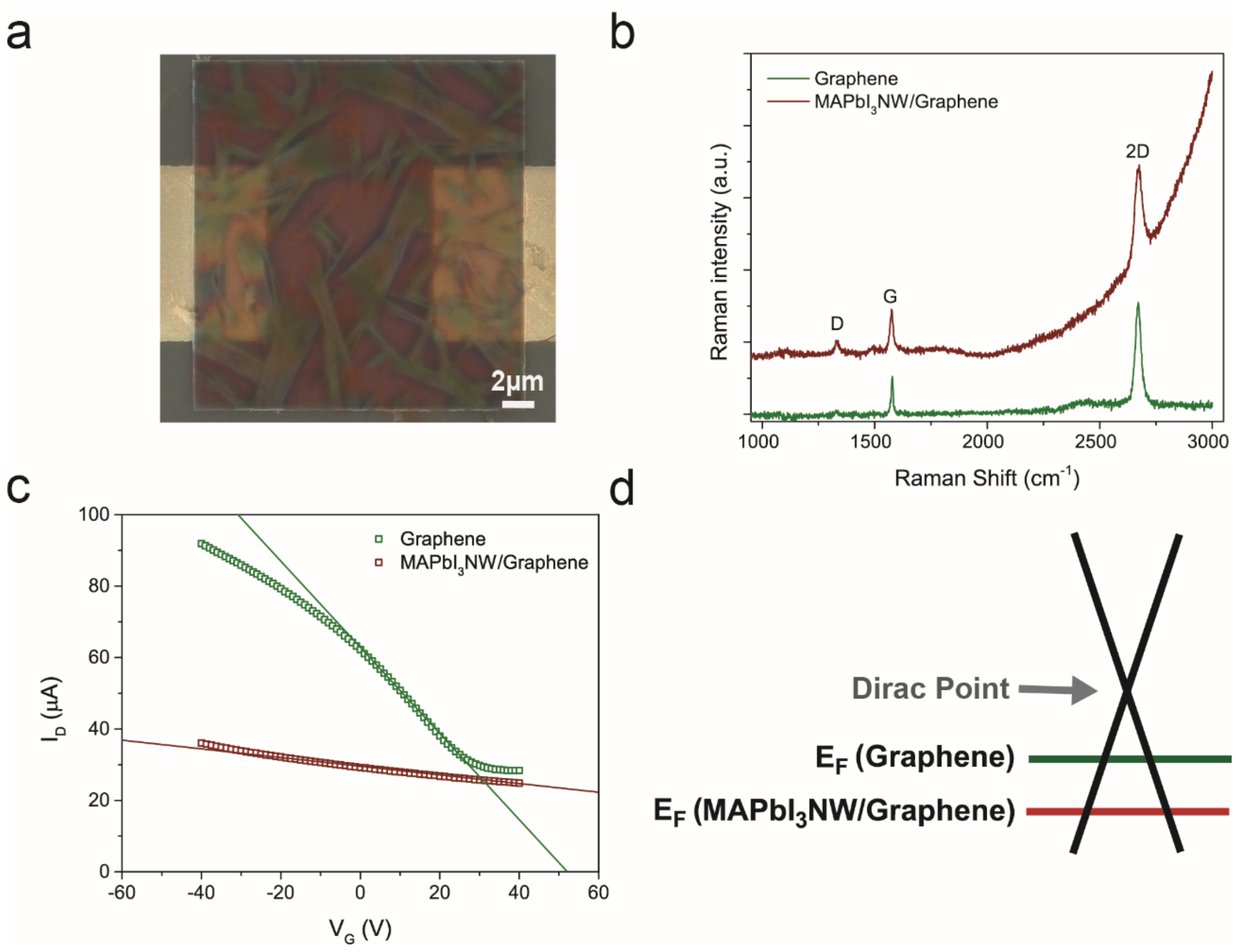

**Figure 1.** (a) Combined Optical and SEM micrograph of a representative device. (b) Raman spectra of the graphene before (green line) and after (red line) the deposition of the $MAPbI_3$ nanowires (spectra are vertically shifted for clarity. (c) Transfer characteristic of a representative device without illumination, before (green line) and after (red line) the deposition of the network of $MAPbI_3$ nanowires. (d) Dirac cone scheme of the graphene before (green line) and after (red line) the hybridization.

The device photoelectrical response was tested under broad-band illumination conditions using a halogen lamp with intensities from 3.3 pW to 212.5 pW (schematic representation shown in figure 2a). Under light illumination, electron-hole pairs were generated in the $MAPbI_3$ nanowires and separated by the internal electric field due to the work function mismatch between graphene and the perovskite nanowire network; analogous to what was reported by Konstantatos *et al.* for the graphene/PbS junction[18]. Positive charge carriers were injected into the graphene while the negative charges accumulated in the nanowires acted as an additional light tunable gate that further decreased the Fermi energy of graphene[4, 19] (Figure 2b).

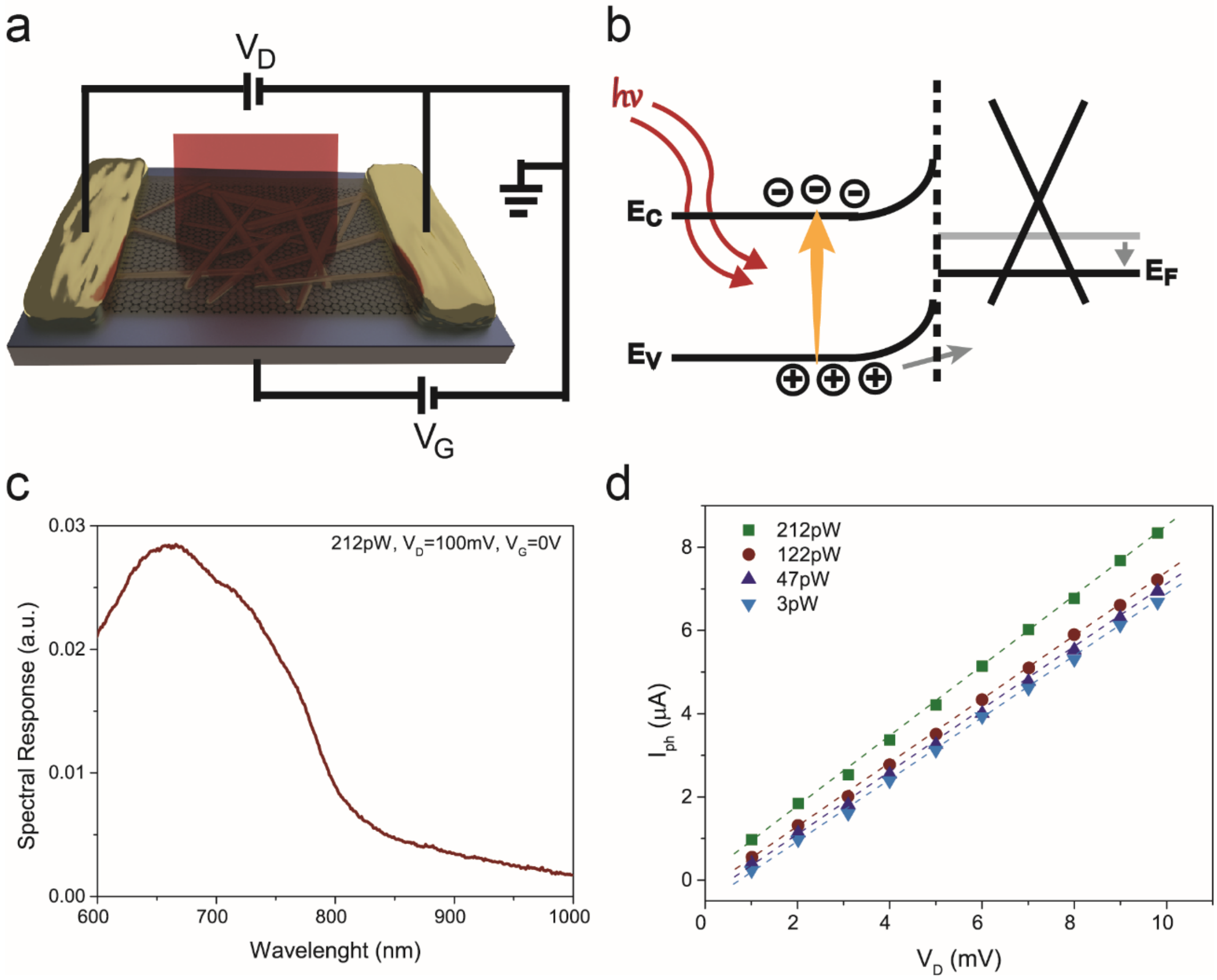

**Figure 2.** (a) Schematic representation of the measurement setup. (b) Band diagram of the nanowire/graphene heterojunction for gate voltages ($V_G$) lower than the charge neutrality point. The injection of the photogenerated holes into the graphene is facilitated by the built-in field formed at the junction. The electrons accumulated in the perovskite nanowires generate an electric field that modulates further the carriers in the graphene. (c) Variation of the current in the device as a function of the wavelength of the incident light. The sharp increase below ~770 nm *i.e.* below the band gap of $MAPbI_3NW$ testifies that the photosensitivity is due to the $MAPbI_3NW$ network. (d) Output characteristic of the photodetector for different incident light intensities. The linear behavior and absence of saturation indicates the possibility that higher responsivity can be reached increasing the voltage between the two metal contacts.

Wavelength-dependent photocurrent measurements revealed a central role of $MAPbI_3$NWs in the photon induced carrier generation (Figure 2c). The photocurrent generation is observed in the wavelength range characteristic for the band structure (onset ~770 nm, ~1.6 eV) of $MAPbI_3$. The number of photo-induced charge carriers increases proportionally with the applied drain voltage without the sign of any saturation at all measured incident light intensities (Figure 2d). We estimate the responsivity of the hybrid device to be as high as $2.6x10^6$ A/W at $V_D$=10 mV and 3 pW. According to the authors' knowledge, at present, this champion device has the highest responsivity values amongst the perovskite based photodetectors. Considerably, the observed value is 7 orders of magnitude higher than that of the commercially available Si detectors (0.6 A/W) and 4 orders of magnitude higher than the responsivity of hybrid graphene/$MAPbI_3$ nanoparticles photodetector reported very recently.[7] We attribute this enormous enhancement of responsivity to engineering the contact between the $MAPbI_3$NW sensitizer and the graphene electrode by synthetizing $MAPbI_3$ nanowire network.

Now we turn to the discussion of the limiting factors of the responsivity and comment on further possibilities to improve the responsivity of $MAPbI_3$ nanowire/graphene photodetectors. As posited, the charge collection efficiency is one of the most critical parameters in photodetectors. Overall, the recombination losses can be often directly correlated to the transit time, hence the channel length of the device and the electric field used to extract the photogenerated charges from the device. Here, we tested the photoelectrical response of about 20 devices with different channel lengths (ranging from 10 μm to 120 μm) under a 633 nm red laser illumination of 2.5 $Wcm^{-2}$. For a fixed source-to-drain electric field of 8 V/cm, the photoresponse of the devices were about 10 times higher when its dimensions were 5 times smaller (Figure 3a). We attribute this 10-fold increase in responsivity to a more effective collection of the photogenerated charge carriers in the device. The scaling law indicates the crucial role of $MAPbI_3$ nanowire/graphene interface quality and demonstrates that miniaturization is a way forward to further improve the photodetector performance.

The undeniable role of the quality of $MAPbI_3$NW/graphene interface is also indicated by the time dependence of the photocurrent. The time response showed characteristic rise and fall times of ~55 s and ~75 s of the photocurrent, respectively (Figure 3b). Even though our perovskite nanowire based hybrid photodetector is faster than other highly efficient graphene-based photodetectors[13], the measured response time values are still much longer than the instrinsic photoresponse time of the methylammonium lead iodide perovskite[20] and other perovskite-based photodetectors[4, 7, 14]. As may be observed in Figure 3b the response times can be divided into two components. The fast component has characteristic time scales of ~5 s (corresponding to ~70% decay) and the slow component attributed to the multiplicity of charge traps in the nanowire film stems from different surface states[21]. It is reasonable to suppose that the improvement of the interface between the graphene and the nanowires and the reduction of the defect density states could significantly increase the responsivity, decrease the trapped electron lifetime, hence, reduce the charactereistic response times from tens of seconds down to a few milliseconds or shorter.

Currently little is known about the physicochemical properties of the interfaces of organolead halide perovskites and carbon nanostructures such as fullerenes, graphene and carbon nanotubes. In solid state perovskite based solar cells, integrated nanomaterials of different dimensionalities serve a beneficial role on the photovoltaic device stability (i.e. reduced I-V hysteresis, non-corrosive effect, efficient charge collector and moisture protection), as recently demonstrated by several research groups. Nevertheless, little is known about the physicochemical properties of these interfaces and their role in the charge transfer process. In order to test the $MAPbI_3$NW/graphene electronic junction we have fabricated two terminal longitudinal devices with perovskite nanowire active layer and two graphene contacts (Figure

S3). These devices showed linear I-V characteristics proving that there is a non-rectifying ohmic type contact between the CVD graphene and the $MAPbI_3NW$ (Figure S3). This result is somewhat surprising, since it is well known that *e.g.* in circuit fabrication generating low resistance ohmic contact allowing the photo generated charges to flow easily in both directions between a metal and a semiconductor usually requires careful techniques. The strong reducing/oxidizing power of the highly concentrated ionic solution of methylammonium lead iodide in organic solvents brought in contact with graphene during the slip-coating process has twofold role in the ohmic contact formation. First, it acts as the source of perovskite crystallization. Second, during the nanowire growth it creates a highly reducing environment (similar to high vacuum cleaning and thermal annealing) and acts as a local "etchant" by removing, dissolving the adsorbed impurities. The perovskite nanowires then start to grow on the graphene surface by classical supersaturation induced nucleation and crystal growth. Importantly, the anisotropic crystal growth is guided by the organic solvent itself trough and intermediate solvatomorph formation and subsequent solvent evaporation induced perovskite crystallization, ultimately resulting in the formation of low resistance ohmic contacts between the two materials. Currently we are investigating the redox properties of methylammonium lead iodide solutions more in detail. Conclusive data will be reported later.

As it is depicted in Figure 3d, in the $MAPbI_3NW$/graphene transistor the application of a transversal electrical field (induced by the back-gate voltage $V_G$) modulates the photoresponsivity by tuning the graphene conducting channel. Application of $V_G$=50 V increased the responsivity by 15% while inverting the voltage reduces the responsivity by about the same amount. Indicating that reducing the shift of the graphene neutrality point enhances the responsivity. Future efforts, optimization and device engineering could result in more robust back-gate dielectrics sustaining larger electric fields. Overall, positive effect of augmented $V_G$ on the response time is also expected[18].

The photogating effect, the shift in the charge neutrality point of the graphene upon light illumination to the $MAPbI_3$NW/graphene interface is shown in figure 3d inset. We estimated this shift in by relating the photoinduced current variation to the values that the back-gate voltage should have to induce the same variation. As a consequence of the generated photocurrent, the shift of the Dirac point increases extremely fast for low intensities (figure 3d inset). This is in good agreement with the observed reduction of the responsivity at high light-intensities i.e. when the photogating effect is high.

The photocurrent shows no saturation as a function of the drain voltage (Figure 2d) suggesting that higher responsivities could be readily achieved by applying a larger longitudinal electric field[22]. Our hybrid device acts as a solid state photomultiplier converting 1 incoming photon (at 633 nm) to ~$4 \times 10^6$ outgoing electrons. This suggests the potential application of this device as a room temperature single photon detector as 1 photon/sec illumination would generate 0.6 pA currents what is within the detection limit of commercial high-sensitivity current meters. Furthermore, these values were obtained at very low longitudinal electric fields (~10 V/cm) meaning that single electron detection could be achieved even with low power consumption.

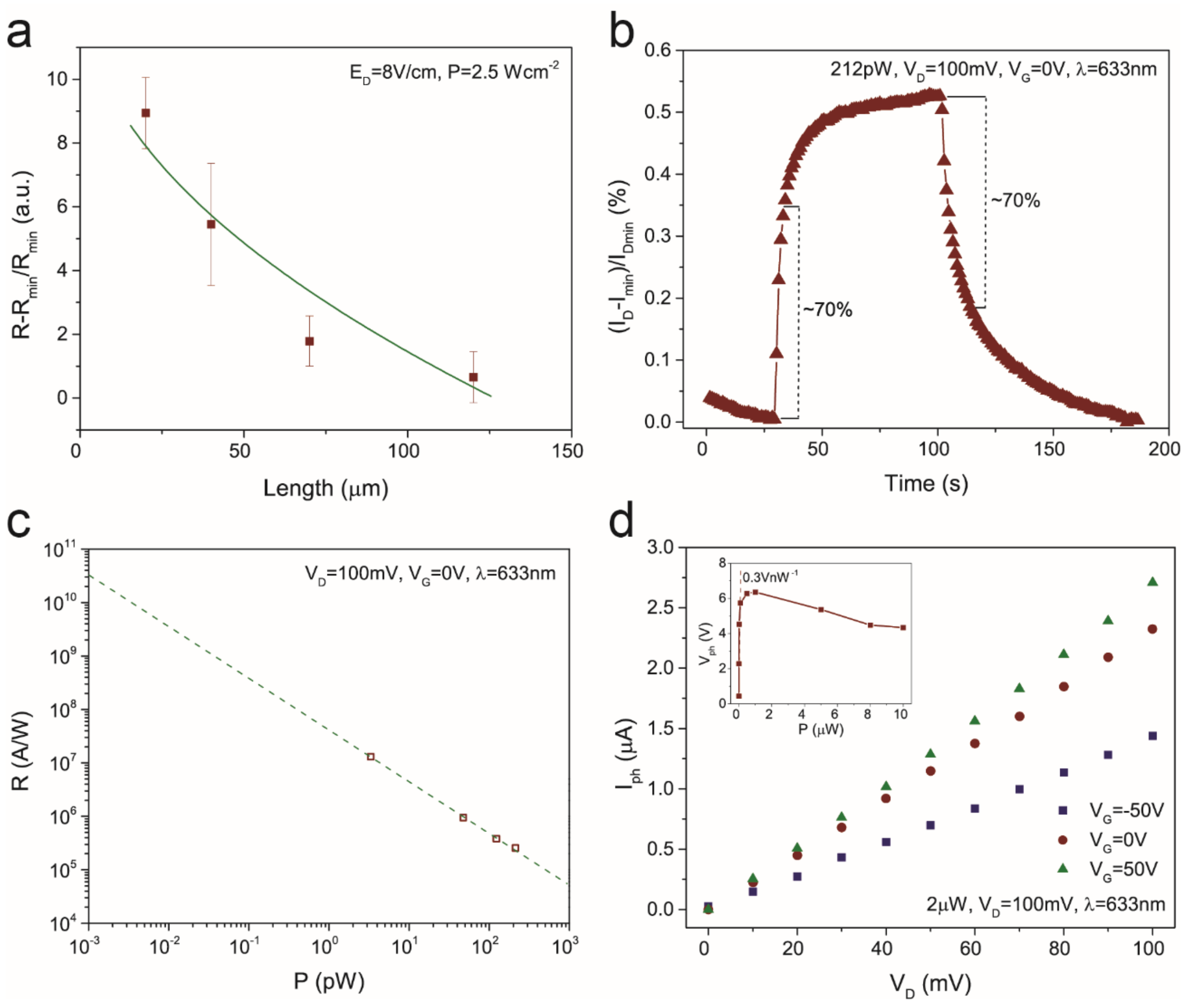

**Figure 3.** (a) Relative variation of the responsivity as a function of the device length for a fixed source-to-drain electric field E=0.83 Vcm$^{-1}$. (b) Time response of a representative device. Two regimes can be identified: a fast one corresponding to ~70% decay (~5 s) and a slow one of ~75 s associated with the charge traps in nanowire film. (c) Responsivity of the device for different incident light intensities. Dashed line show extrapolation towards lower intensities. (d) Photocurrent of the bottom-gated device as a function of both the applied bias and gate voltage. The performances of the device increase with both voltages. Inset presents the photogating $V_{PH}$ as a function of light intensity. $V_{PH}$ raises rapidly with increasing light intensity and saturates around 1 μW.

Finally the photoresponse of the detector dramatically increases at very low light intensities. The maximum responsivity of ~2.6x10$^6$ A/W was measured with an incident power of 3.3 pW (Figure 3c). Assuming the observed relation between the photocurrent and the incident power (Figure 3c) is valid even at lower power intensities, i.e. in the fW range, efficiencies of the order of $10^{10}$ A/W could be achieved at room temperature.

In conclusion, we fabricated the first hybrid $MAPbI_3$ nanowire/graphene photodetector. The measured device photoresponsivity was as high as $2.6x10^6$ A/W that is 4 orders of magnitude higher the best hybrid perovskite photodetector reported by Y. Lee *et al.*[7] and comparable to the best hybrid graphene photodetectors reported so far[18,13] under similar operating conditions. We attribute these very high device performances mainly to the nanowire perovskite morphology. The drastic enhancement of the responsivity at very low light intensities (pW) suggest the use of $MAPbI_3$ nanowire/graphene devices as low-light imaging sensors and single photon detectors.

# *Supporting Information*

# *Micro-engineered $CH_3NH_3PbI_3$ nanowire/graphene phototransistor for low intensity light detection at room temperature*

*M. Spina, M. Lehmann, B. Náfrádi , L. Bernard, E. Bonvin, R. Gaál, A. Magrez, L. Forró, E. Horváth**

Laboratory of Physics of Complex Matter (LPMC), Ecole Polytechnique Fédérale de Lausanne, 1015 Lausanne, Switzerland

**Methods**

*Graphene growth*

The graphene used in this work was grown on copper foils via a Chemical Vapor Deposition (CVD) method. First, a 25µm-thick copper foil with lateral dimensions of 2.2 x 5.5 $cm^2$ (Alfa Caesar, 99.8%) was inserted into a 2.5cm-large quartz tube, then placed inside a horizontal tube furnace. After the quartz tube was evacuated to a pressure of 4 mTorr, high purity hydrogen (Carbagas $H_2$, purity 60) was introduced to reach a pressure of 3 Torr. Then, the system was heated to 1000 °C at a rate of 50°C/min (pressure ~ 900 mTorr). The copper foils were annealed at the growth temperature (1000 °C) in hydrogen for 30 min. A pulsed flow of methane was then introduced to grow the graphene for 3 min. After the growth, the system was cooled down to 300 °C at a rate of 6 °C/min in a hydrogen atmosphere. Samples were removed from the chamber at temperature below 300 °C.

*Graphene characterization*

To investigate the quality of the CVD-grown graphene, Raman spectroscopy was performed with a confocal Raman spectrometer (LabRam HR, Horiba Scientific) using a 532nm-laser and a 100x objective lens (spot size of 0.3 µm). From the spectra shown in figure S1, we can observe that the G and 2D band point are at 1578 and 2670 $cm^{-1}$ respectively and have a full width at half maximum of 15 and 33 $cm^{-1}$. After the sensitization of the graphene, quantitative analysis of the G and 2D peaks resulted difficult because of the intense background signal coming from the photoluminescence of the $MAPbI_3$ nanowires.

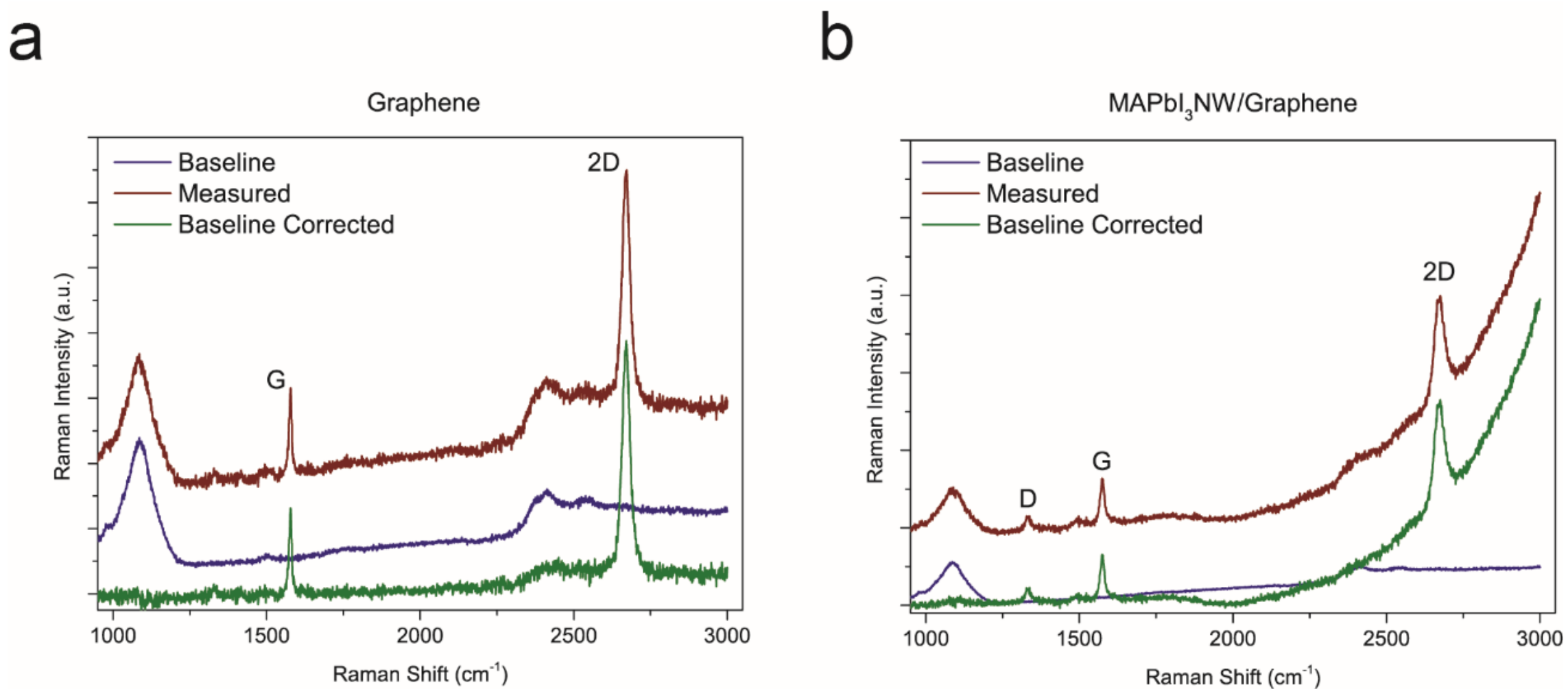

**Figure S1:** Raman spectra of the graphene film (a) and $MAPbI_3$NW/graphene junction (b) used in the study.

*Device Fabrication*

The graphene was transferred onto a highly p-doped Si substrate with a 300nm thick $SiO_2$ thermally grown on top using the stamp method[1]. First, one side of the Cu/graphene film was spin-coated (1000 rpm, 1 min) with a layer of poly(methyl methacrylate) (PMMA, MW 495K). Then, the copper foil was cured 5 min at 180 °C and the non-coated graphene layer was removed using sand paper. The sample was then dipped in an iron chloride solution (20 mg/ml) for 7-8 hours in order to etch the copper foil. Later, the graphene/PMMA film was rinsed in a bath of deionized water and in a diluted hydrochloric acid solution to remove the remaining ions. Finally, the graphene/PMMA film was picked up by the substrate, and dried in ambient atmosphere. After the transfer of the graphene, the PMMA layer used in the process was patterned by e-beam lithography, and employed as an etch mask for patterning the graphene by a directional plasma oxygen (100W, 10s, 20sccm $O_2$). The resist was then removed by immersion of the sample in acetone and rinse in Isopropyl alcohol (figure S2a). 100nm-thick gold contacts were then patterned by e-beam lithography and thermally evaporated on top of the graphene film (figure S2b). The network of $MAPbI_3$ nanowires was subsequently deposited by the slip- coating technique reported by Horváth et al.[2] (figure S2c). The devices were then encapsulated in a PMMA layer in order to avoid degradation of the nanowires because of exposure to the ambient atmosphere (figure S2d). The same PMMA layer was patterned by e-beam and used as an etch mask to remove the nanowires located outside the device area by a directional Ar ion etching (figure S2e). A representative schema of the final device is show in figure S2f.

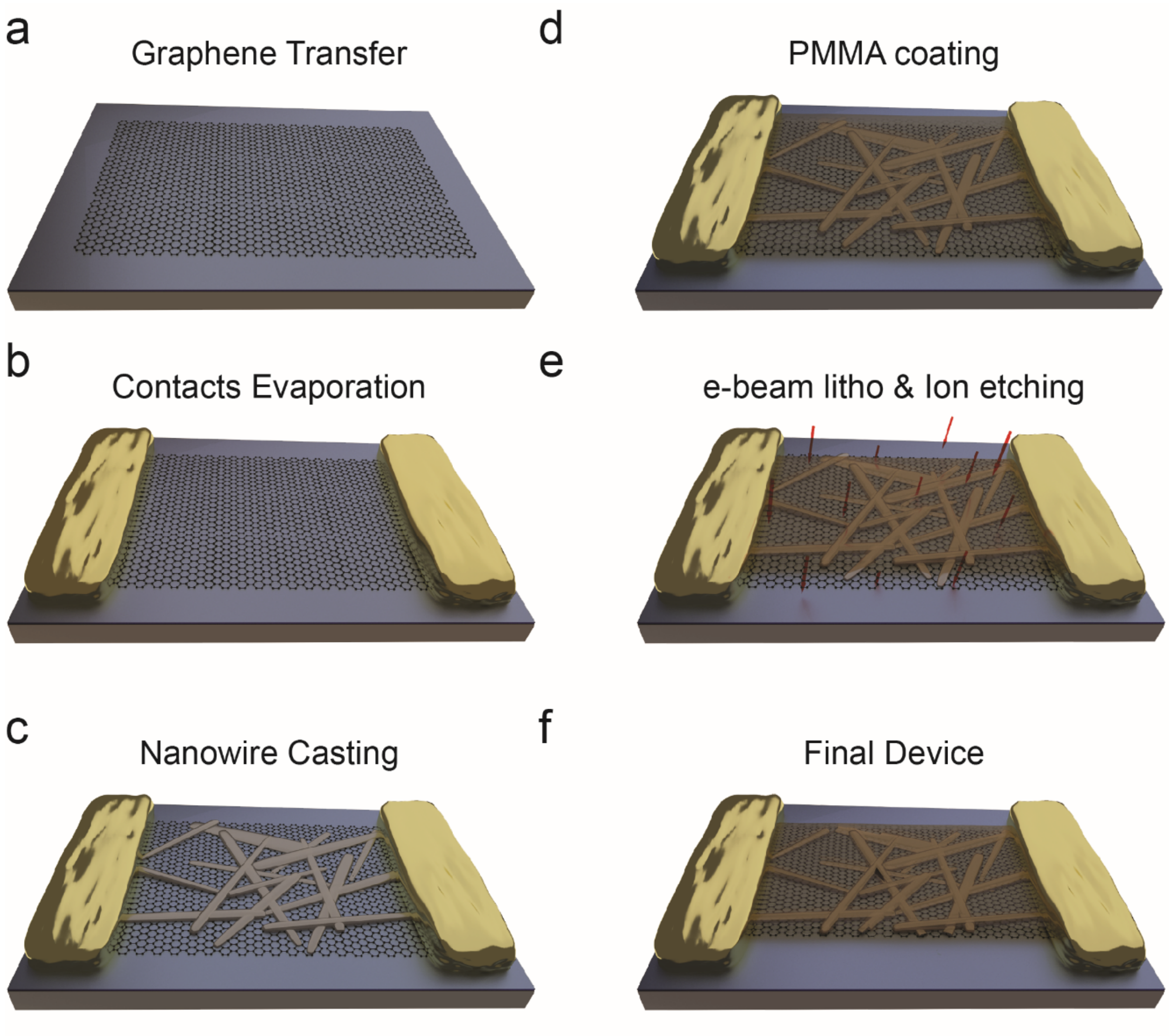

**Figure S2**. (a) The graphene film is transferred on a highly p-doped Si substrate with 300nm $SiO_2$ thermally grown on top. (b) Au contacts are patterned by e-beam lithography and thermally evaporated on the substrate. (c) $MAPbI_3$ nanowires are slip-coated on the patterned graphene film. (d) A film of PMMA is spin coated on the device and used as etch mask. (e) The nanowires outside the device are etched away by Ar ions. (f) Final device.

*Photoelectrical characterization*

The photoelectric response measurements of the fabricated hybrid devices were performed using a standard DC technique. The light sources used were a red laser beam (λ=633 nm) with a spot size of about 4 mm and a halogen lamp connected to a monochromator for the characterization under broad-band illumination conditions. All the measurements were performed at room temperature and in ambient environment.

*Study of the graphene-perovskite interface*

In photovoltaic devices as for every optoelectronic devices, proper contact between electrode and p/n type material is essential for having an efficient charge collection. To investigate the type of contact between the graphene and the perovskite nanowires we fabricated a photodetector were a network of interconnected $CH_3NH_3PbI_3$ nanowires is contacted by two foils of monolayer graphene acting as the source and drain electrode (figure S3a).

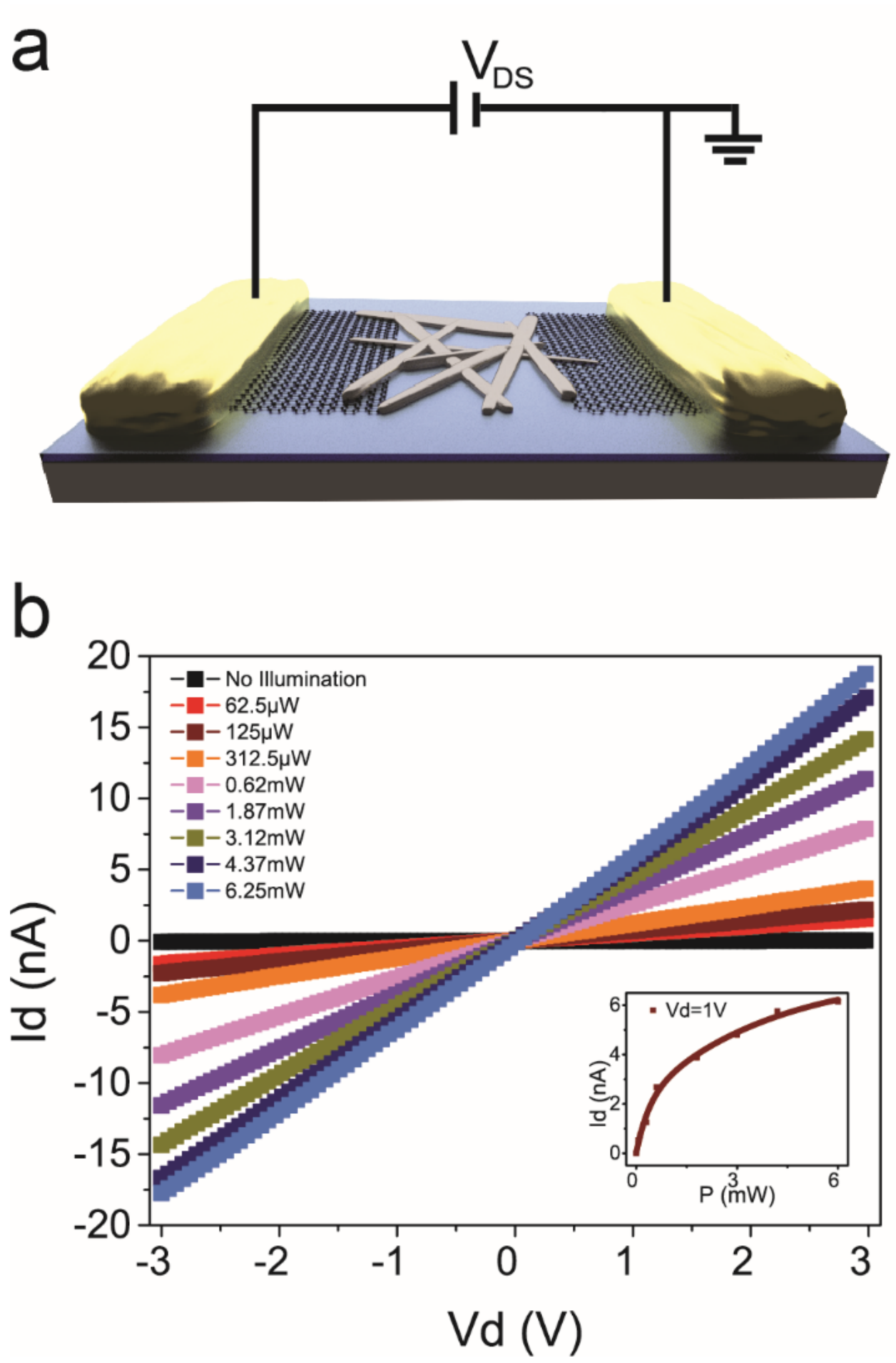

**Figure S3**. (a) Schematic representation of the graphene-nanowires-graphene device. (b) Output characteristics for different power intensities. Graphene contacts show non-rectifying ohmic behavior. The drain current tends to saturate for high laser power intensities but saturation has not been achieved in this study.

The current in the device was measured as a function of the applied source-to-drain voltage in the dark and under illumination of a red laser (λ=633 nm) with a power density of 2.5 $Wcm^{-2}$ (Fig 3b). The output characteristics follow a linear behavior, indicating that the contacts between the nanowires and the graphene films are ohmic. In the dark state, the device behaves like a good insulator with currents of the order of tens of pA and resistances in the GOhm range. Under the illumination of the laser, the absorption of the light generates electron-hole pairs that are extracted by the source-to-drain electric field and cause an increase in the conductance of the material up to a factor of 500. We probed the photoresponse of the device under different incident laser powers in the 60 μW to 6 mW range and observed a parabolic increase of the current with the incident power intensity (inset to Fig S3b).

*Spectral responsivity measurement*

Photocurrent as a function of incident light energy was measured in a two-terminal geometry with fixed bias voltage. For photocurrent spectra a low intensity monochromatic light was selected by a MicroHR grid monochromator from a halogen lamp. The wavelength resolution (FWFM) of the 600 gr/mm grating was 10 nm.

*Estimation of photomultiplication*

For our device presented in Figure 2d 3 pW 633 nm light illumination generates 6 μA photocurrent. A single photon energy at 633 nm is $3.138\ 144\ 689\text{x}10^{-19}$ J thus 3 pW illumination 9559789 phonons/second. 6 μA current is a flow of $3.7446\text{x}10^{13}$ electron/sec by definition. Accordingly 1 incoming phonon is transformed to ~$4\text{x}10^{6}$ electrons in the device. This value is equivalent to 0.6 pA current in 1 photon/sec illumination conditions.

*Comparative table of the performances of the best-in-class graphene and* $MAPbI_3$ *photodetectors*

| Active Material | Spectral Window (nm) | Incident Power | Responsivity (A/W) | $V_G$ (V) | $V_{SD}$ (V) | Reference |
|---|---|---|---|---|---|---|
| ME-SLG/ PbS QDs | 532 | 8fW | ~$5\text{x}10^{7}$ | -20 | 5 | G. Konstantatos, et al.[3] |
| ME-SLG/ ME-MLMo$S_2$ | 635 | 6.4fW$\mu m^{-2}$ ~320fW | ~$5x10^8$ | -50 | 0.1 | K. Roy, et al.[4] |
| CVD-SLG/ MAPb$I_3$ NPs | 520 | 1μW | 180 | 0 | 0.1 | Y. Lee, et al.[5] |
| MAPb$I_3$ NPs | 365 | 10 $\mu Wcm^{-2}$ | 3.49 | 0 | 3 | X. Hu et al.[6] |
| MAPb$I_3$ NWs | 633 | 340nW | $5x10^{-3}$ | 0 | 1 | E. Horváth et al.[2] |
| ME-SLG | 532 | - | $8.6x10^{-3}$ | 0 | 0.1 | Y. Zhang et al.[7] |
| CVD-SLG/ MAPb$I_3$ NWs | 633 | ~3 pW | ~$2x10^6$ | 0 | 0.01 | This work |